\documentclass[letterpaper,12pt]{article}
\pdfoutput=1 
\usepackage{jheppub}

\usepackage{graphicx}

\usepackage{amsmath}
\usepackage{array}

\DeclareMathOperator{\Gr}{G}

\title{
Symbol Alphabets from Plabic Graphs III: $n=9$
}

\author[1]{J.~Mago,}
\author[2]{A.~Schreiber,}
\author[1,3]{M.~Spradlin,}
\author[1]{A.~Yelleshpur Srikant}
\author[1]{and A.~Volovich}

\affiliation[1]{Department of Physics,
Brown University,
Providence, RI 02912, USA}

\affiliation[2]{The Mathematical Institute,
University of Oxford,
Woodstock Road,
OX2 6GG, UK}

\affiliation[3]{Brown Theoretical Physics Center,
Brown University,
Providence, RI 02912, USA}

\abstract{Symbol alphabets of $n$-particle amplitudes in $\mathcal{N}=4$ super-Yang-Mills theory are known to contain certain cluster variables of $\Gr(4,n)$ as well as certain algebraic functions of cluster variables.
In this paper we solve the $C\,Z = 0$ matrix equations associated to
several cells of the totally non-negative Grassmannian,
combining methods of
arXiv:2012.15812 for rational letters and arXiv:2007.00646 for
algebraic letters.
We identify sets of parameterizations of the top cell of $\Gr_+(5,9)$ for which the
solutions produce all of (and only) the cluster variable letters
of the 2-loop nine-particle NMHV amplitude, and identify plabic graphs
from which all of its algebraic letters originate.
}

\begin{document} 

\maketitle

\section{Introduction}

The fact that cluster algebras~\cite{FZ1,FZ2,Scott,FWZ} govern the symbol
alphabets~\cite{Goncharov:2010jf} of multiloop $n$-particle amplitudes in planar maximally
supersymmetric Yang-Mills (SYM) theory is by now well-established for
$n=6,7$~\cite{Golden:2013xva}
(see~\cite{Caron-Huot:2020bkp} for a review of recent progress on
the computation of these amplitudes via bootstrap).
Starting at $n=8$ qualitatively new features arise, which have been
studied via several different approaches (see for
example~\cite{Arkani-Hamed:2019rds,Drummond:2019cxm,Henke:2019hve,
Mago:2020kmp,He:2020uhb,Mago:2020nuv,Herderschee:2021dez,RSV,HP}).

In this paper we continue the program outlined
in~\cite{Mago:2020kmp,He:2020uhb,Mago:2020nuv}, which is based on
the observation that symbol letters of SYM theory seem to naturally emerge
from certain plabic graphs~\cite{Arkani-Hamed:2016byb}
(or equivalently, Yangian invariants).
Specifically, if $Z$ is an $n \times 4$ momentum twistor matrix
parameterizing the kinematic data for an $n$-particle scattering
process, and if $C$ is a $k \times n$ matrix parameterizing a
$4k$-dimensional cell of the totally non-negative
Grassmannian $\Gr_+(k,n)$~\cite{Postnikov},
then solving the matrix
equations $C Z = 0$~\cite{ArkaniHamed:2009dn,Mason:2009qx} sets the parameters
of $C$ to various rational or algebraic functions of
Pl\"ucker coordinates on $\Gr(4,n)$ that often turn out to be
products of symbol letters of amplitudes.

In~\cite{Mago:2020kmp,He:2020uhb} an example for $(k,n) = (2,8)$
was considered that precisely reproduces all of the 18
algebraic symbol letters known to appear in the 2-loop eight-particle
NMHV amplitude~\cite{Zhang:2019vnm}.  At the same time it was
pointed out that if the cell parameterized by $C$ is not the
top cell (i.e., the one with dimension $k(n{-}k)$), then one
generally encounters rational quantities that are not
expressible in terms of cluster variables.
On the other hand, in~\cite{Mago:2020nuv} it was shown that
for any cluster parameterization of the top cell (not necessarily
one associated to a plabic graph), this procedure will \emph{only}
give cluster variables.

Here our focus is on the case $n=9$, where the most up-to-date
symbol alphabet information comes from the computation of the
two-loop NMHV amplitude~\cite{He:2020vob}.  We show how to obtain
all known $n=9$ symbol letters from cluster parameterizations of
cells of $\Gr_+(k,9)$.
First, we provide an explicit list of cluster parameterizations of the
top cell of $\Gr_+(5,9)$ which collectively provide all 531
of the $n=9$ rational letters found in~\cite{He:2020vob} (and no
additional letters).  Second, we identify a cyclic class of
parameterizations of cells
of $\Gr_+(3,9)$ which
collectively provide all 99 of the $n=9$ algebraic letters, together with
a few additional algebraic quantities.

As already acknowledged
in~\cite{Mago:2020kmp,Mago:2020nuv}, we do not as of yet have
a ``theory'' to explain the pattern of which cells are associated
to cluster variables (or algebraic functions thereof) that are
actually observed to appear in amplitudes.  Instead, we view
our work as providing some kind of ``phenomenological'' data in
the hope that future work will be able to shed more light on
this interesting problem.

\section{Rational Letters}

\subsection{\texorpdfstring{$n=8$ Extended Rational Alphabet}{n=8 Extended Rational Alphabet}}

To date, a total of 180 rational letters, all of which are cluster variables of
$\Gr(4,8)$, are known to appear in the eight-particle amplitudes of SYM theory.
These letters are tabulated in~\cite{Zhang:2019vnm}.
By studying a certain fan one can naturally associate to the tropical positive Grassmannian
(or, equivalently, its dual polytope), \cite{Arkani-Hamed:2019rds,Drummond:2019cxm,Henke:2019hve}
encountered a larger list of cluster variables that includes these 180, together with 100 more.  These additional variables may
appear in the symbols of eight-point amplitudes that have
not yet been computed.
We call this collection of 280 cluster variables the $n=8$ \emph{extended rational alphabet};
it consists of
\begin{itemize}
\item 68 four-brackets of the form $\langle a ~ a{+}1 ~ b  ~ c \rangle$, 

\item 8 cyclic images of $\langle 1 2 \bar{4} \cap \bar{7} \rangle$, 

\item 40 cyclic images of $\langle 1 (23) (45)(78) \rangle$, $\langle 1 (23) (56) (78) \rangle$, $\langle 1 (28) (34) (56) \rangle$, $\langle 1 (28) (34) (67) \rangle$, $\langle 1 (28) (45) (67) \rangle$,

\item 48 dihedral images of $\langle 1 (23) (45) (67) \rangle$, $\langle 1 (23) (45) (68) \rangle$, $\langle 1 (28) (34) (57) \rangle$,

\item 8 cyclic images of $\langle \bar{2} \cap (245) \cap \bar{8} \cap (856) \rangle$, 

\item 8 distinct images of $\langle \bar{2} \cap (245) \cap \bar{6} \cap (681) \rangle$, 

\item 16 dihedral images of $\langle \langle 12345678 \rangle \rangle$,
 
\item 2 letters, $\langle 1357 \rangle$ and $\langle 2468 \rangle$,

\item 8 cyclic images of $\langle 1 (23) (46) (78) \rangle$, 

\item 16 dihedral images of $\langle 1 (27) (34) (56) \rangle$,

\item 2 cyclic images of $\langle \bar{2} \cap \bar{4} \cap \bar{6} \cap \bar{8} \rangle$,

\item 8 cyclic images of $\langle \bar{2} \cap (246) \cap \bar{6} \cap \bar{8} \rangle$,

\item 32 dihedral images of $\langle \langle 12435678 \rangle \rangle$, $\langle \langle 12436578 \rangle \rangle$,

\item 16 dihedral images of $\langle 1234 \rangle \langle 1678 \rangle \langle 2456 \rangle - \langle 1267 \rangle \langle 1348 \rangle \langle 2456 \rangle + \langle 1248 \rangle \langle 1267 \rangle \langle 3456 \rangle$,

\end{itemize}
Here $\langle abcd\rangle$ are Pl\"ucker coordinates on $\Gr(4,n)$ and we define 
\begin{align}
\begin{split}
\bar{a} &\equiv (a{-}1~a~a{+}1) \, , \\
\langle a (bc) (de)(fg) \rangle &\equiv \langle abde \rangle \langle ac fg \rangle - \langle acde \rangle \langle abfg \rangle \, ,\\
\langle a,b,c, (de)\cap(fgh) \rangle & \equiv \langle abc d \rangle \langle efgh \rangle - \langle abce \rangle \langle dfgh \rangle \, , \\
\langle x,  y , (abc) \cap (def) \rangle &\equiv \langle x a b c \rangle \langle y d e f \rangle -  \langle y a b c \rangle \langle x d e f \rangle \, , \\
\langle \langle abcdefgh \rangle \rangle &\equiv \langle abcd \rangle \langle abef \rangle \langle degh \rangle - \langle abde \rangle \langle abef \rangle \langle cdgh \rangle \\
&\qquad+ \langle abde \rangle \langle abgh \rangle \langle cdef \rangle\,,\\
\langle \bar{x} \cap (abc) \cap \bar{y} \cap (def) \rangle &\equiv   \langle a , (bc) \cap \bar{x} , d , (ef) \cap \bar{y} \rangle\,.
\end{split}
\end{align}

\begin{figure}[ht]
\centering
\includegraphics[width=0.65\textwidth]{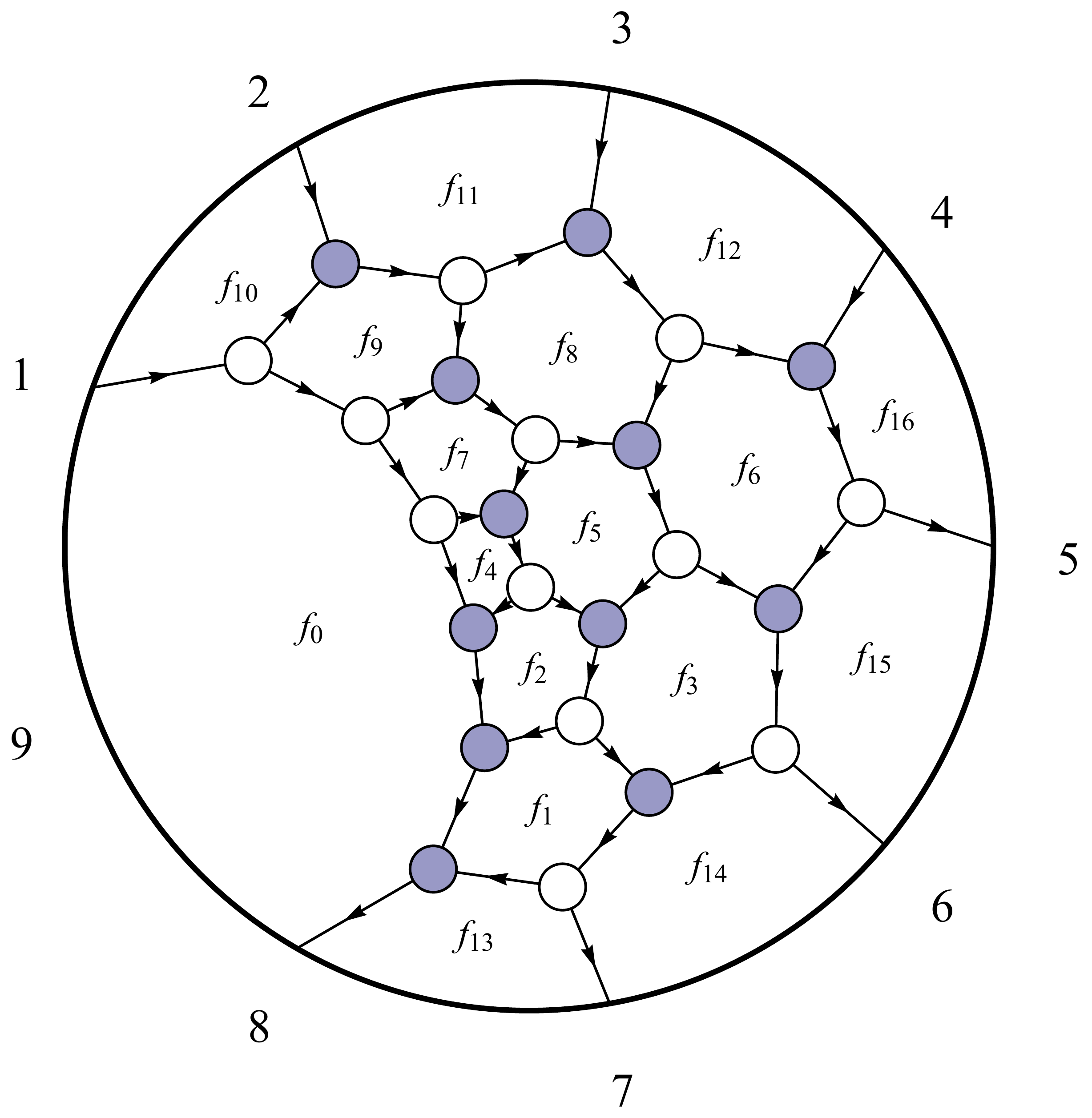}
\caption{A plabic graph associated to the top cell of G${}_+ (4,8)$.}
\label{fig:n=8_initialgraph}
\end{figure}

We know from~\cite{Mago:2020nuv} that for any cluster parameterization $C$
of the top cell of $\Gr_+(4,8)$, solving $C Z = 0$ expresses the parameters of $C$ in terms of products of
powers of $\Gr(4,8)$ cluster variables.  Our aim is to identify a set of parameterizations that
collectively involve precisely the 280 letters of the extended rational alphabet (and no other letters).

We begin by taking $C$ to be the boundary measurement of the plabic graph shown in Fig.~\ref{fig:n=8_initialgraph}
(see~\cite{Mago:2020kmp,Mago:2020nuv} for more details on our conventions).
Then the solution to $C Z = 0$ is given by
\begin{align} \label{n=8sol}
\begin{array}{lll}
f_0 =-\frac{\langle 1234\rangle }{\langle 2348\rangle } \, , ~~ &  
f_1=\frac{\langle 3458\rangle  \langle 4567\rangle }{\langle 3456\rangle  \langle 4578\rangle }\, , ~~ &
f_2=\frac{\langle 2348\rangle  \langle 3456\rangle  \langle 4578\rangle }{\langle 2345\rangle  \langle 3478\rangle  \langle 4568\rangle }  \, , \\ & & \\
f_3=\frac{\langle 3478\rangle  \langle 4568\rangle }{\langle 3458\rangle  \langle 4678\rangle } \, , ~~ & 
f_4=\frac{\langle 1238\rangle  \langle 2345\rangle  \langle 3478\rangle }{\langle 1234\rangle  \langle 2378\rangle  \langle 3458\rangle } \, , ~~ &
f_5=\frac{\langle 2378\rangle  \langle 3458\rangle  \langle 4678\rangle }{\langle 2348\rangle  \langle 3678\rangle  \langle 4578\rangle }  \, , \\ & & \\
f_6=\frac{\langle 3678\rangle  \langle 4578\rangle }{\langle 3478\rangle  \langle 5678\rangle } \, , ~~ &
f_7=\frac{\langle 1278\rangle  \langle 2348\rangle  \langle 3678\rangle }{\langle 1238\rangle  \langle 2678\rangle  \langle 3478\rangle } \, , ~~ &
f_8=\frac{\langle 2678\rangle  \langle 3478\rangle }{\langle 2378\rangle  \langle 4678\rangle }  \, , \\ & & \\
f_9=\frac{\langle 1678\rangle  \langle 2378\rangle }{\langle 1278\rangle  \langle 3678\rangle }\, , ~~ & 
f_{10}=\frac{\langle 2678\rangle }{\langle 1678\rangle } \, , ~~ & 
f_{11}=\frac{\langle 3678\rangle }{\langle 2678\rangle }  \, ,\\ & & \\
f_{12}=\frac{\langle 4678\rangle }{\langle 3678\rangle } \, , ~~ &     
f_{13}=-\frac{\langle 4568\rangle }{\langle 4567\rangle } \, , ~~ &
f_{14}=-\frac{\langle 4578\rangle }{\langle 4568\rangle } \, ,  \\ & & \\
f_{15}=-\frac{\langle 4678\rangle }{\langle 4578\rangle } \, , ~~ &
f_{16}=\frac{\langle 5678\rangle }{\langle 4678\rangle } \, . ~~
\end{array}
\end{align}

By drawing the dual quiver (with arrows clockwise around white vertices and counterclockwise around black vertices) and reading off the adjacency matrix,
we can mutate the face variables according to the cluster $\mathcal{X}$-variable mutation rules~\cite{Gekhtman:2002}
\begin{align}
f_k' = \left\{ 
\begin{array}{l r}
f_k^{-1} &~~ i = k \, ; \\
f_i (1 + f_k^{- \text{sgn} (b_{i,k}) } )^{- b_{i,k}} & ~~ i \ne k \, , 
\end{array}
\right.
\end{align} 
where $b_{i,j}$ is the adjacency matrix of the dual quiver. Under mutations, the adjacency matrix transforms as
\begin{align}
b_{i,j}'  = \left\{ 
\begin{array}{l r}
- b_{i,j} &~~ k \in \{ i,j \} \, ; \\
b_{i,j} & ~~ k \notin \{ i,j \} ~~ \text{and} ~~ b_{i,k}  b_{k,j} \leq 0 \, ; \\
b_{i,j} + |b_{i,k} | b_{k,j} & ~~ k \notin \{ i,j \} ~~ \text{and} ~~ b_{i,k}  b_{k,j} > 0 \, . 
\end{array}
\right.
\end{align}

We perform sequences of mutations on internal faces (external faces are considered frozen) and collect all monomial factors that appear in the mutated face variables. We then find a minimal set of mutation sequences for which the mutated face variables collectively contain the entire 280 letter extended rational alphabet (mod cyclic permutations of external labels), and \emph{only} letters from that alphabet. Note that the cluster algebra associated with the dual quiver of the G${}_+ (4,8)$ top cell is of infinite type, and we only search far enough to find minimal length mutation sequences that suffice to produce the entire 280-letter alphabet.

We find that considering all mutation sequences of up to length 5 is sufficient, and in particular we find 13 clusters that are sufficient to generate the entire 280-letter $n=8$ extended rational alphabet (mod cyclic rotations of external labels). These clusters are obtained from the following 13 mutation sequences:
\begin{align} \label{minmutsn=8extended}
\begin{split}
\{&\{4, 7, 8, 3, 6\}\, , ~  \{5, 7, 9, 8, 2\}\, , ~ \{5, 8, 3, 1, 2\} \, ,~ \{6, 8, 7, 4,  2\} \, ,\\
&\{7, 1, 2, 5, 6\}\, ,~ \{7, 2, 3, 6, 5\}\, , ~ \{7, 4, 2, 3, 6\} \, ,~ \{7, 5, 6, 2, 1\} \, , \\
&\{8, 3, 5, 2, 4\}\, ,~ \{8, 4, 5, 1, 3\}\, ,~ \{8, 6, 3, 2, 4\}\, ,~ \{9, 1, 2, 5, 7\}\, , ~ \{9, 8, 5, 3, 1\} \} \, ,
\end{split}
\end{align}
where the sequences should be read as: $\{ a, b, c, \ldots\}:$ mutate on the node $f_a$, then mutate on $f_b$, and then mutate on $f_c$, etc.  It is important to emphasize that this set of minimal length mutational sequences is not unique.  Also, note that at intermediate steps between the initial cluster and the final 13 clusters obtained at the end of these sequences, one can encounter additional cluster variables not contained in the 280-letter alphabet.

\subsection{\texorpdfstring{$n=9$ Rational Alphabet}{n=9 Rational Alphabet}}

To date, a total of 531 rational letters, all of which are cluster variables of $\Gr(4,9)$,
are known to appear in the nine-particle amplitudes of SYM theory.
These letters are tabulated in~\cite{He:2020vob} and consist of:
\begin{itemize}
\item 13 cyclic classes of $\langle 12 kl \rangle$ for $3 \leq k < l \leq 8$ but $(k, l ) \ne (6,7), (7,8)$; 
\item 7 cyclic classes of $\langle 12 (i j k)\cap (lmn) \rangle$ for $3 \leq i < j < k < l < m < n \leq 9$; 
\item 8 cyclic classes of $\langle \bar{2} \cap (245) \cap \bar{6} \cap (691) \rangle$, $\langle \bar{2} \cap (346) \cap \bar{6} \cap (892) \rangle$, $\langle \bar{2} \cap (346) \cap \bar{2} \cap (782) \rangle$, $\langle \bar{2} \cap (245) \cap \bar{7} \cap (791) \rangle$, $\langle \bar{2} \cap (245) \cap (568) \cap \bar{8} \rangle$, $\langle \bar{2} \cap (245) \cap (569) \cap \bar{9} \rangle$, $\langle \bar{2} \cap (245) \cap (679) \cap \bar{9} \rangle$, $\langle \bar{2} \cap (256) \cap (679) \cap \bar{9} \rangle$; 
\item 10 cyclic classes of $\langle 1 (i \, i{+}1) (j \, j{+}1) (k \, k{+}1) \rangle$ for $2 \leq i$, $i+1 < j$, $j+1 < k \leq 8$; 
\item 6 cyclic classes $\langle 1 (2 i) (j \, j{+}1) (k 9) \rangle$ for $3 \leq i < j$, $j+1 < k \leq 8$, but $(i,k) \ne (3,8), (4,7)$; 
\item 14 cyclic classes of $\langle 1 (29) (ij) (k \, k{+}1) \rangle$ for $3 < i < j \leq 8$, $3 \leq k \leq i-2$ or $j+1 \leq k \leq 7$;
\item 1 cyclic class of $\langle 1 , (56) \cap \bar{3} , (78) \cap \bar{3} , 9 \rangle$.
\end{itemize}

\begin{figure}[ht]
\centering
\includegraphics[width=0.65\textwidth]{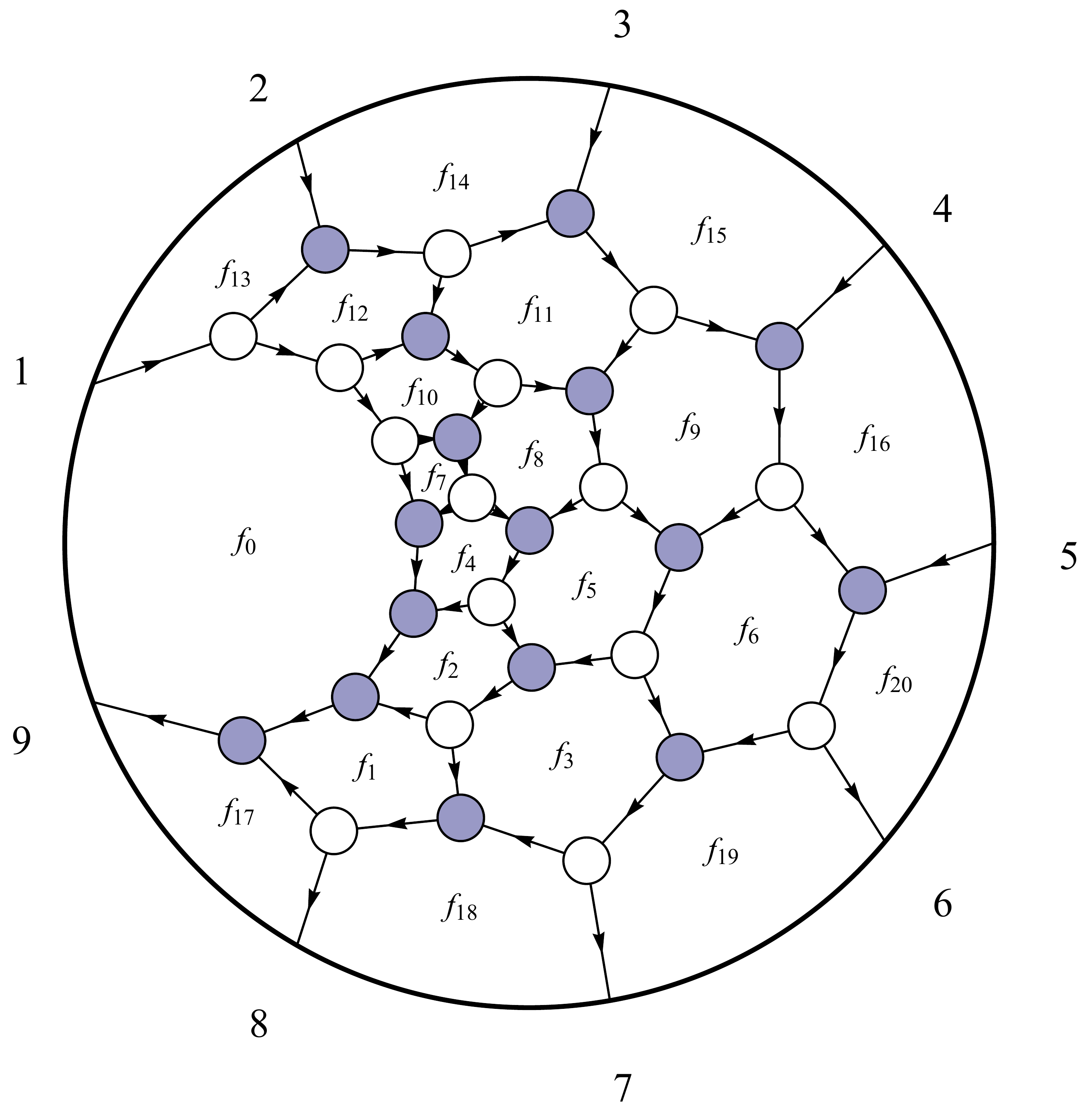}
\caption{A plabic graph associated to the top cell of G${}_+ (5,9)$.}
\label{fig:n=9_initialgraph}
\end{figure}

In this section, we derive this alphabet from plabic and non-plabic parameterizations of the top cell in G${}_+ (5,9)$ by analyzing sequences of mutations on the dual quiver to the initial plabic parameterizations of the top cell,
corresponding to the plabic graph shown in Fig.~\ref{fig:n=9_initialgraph}.
Taking $C$ to be the boundary measurement of this graph, we find that the solution to
$C Z = 0$ is given by
\begin{align} \label{czsol}
\begin{array}{lll}
 f_0=-\frac{\langle 1234\rangle }{\langle 2349\rangle } \, , ~~ &
 f_1=\frac{\langle 4569\rangle  \langle 5678\rangle }{\langle 4567\rangle  \langle 5689\rangle } \, , ~~ &
 f_2=\frac{\langle 3459\rangle  \langle 4567\rangle  \langle 5689\rangle }{\langle 3456\rangle  \langle 4589\rangle  \langle 5679\rangle } \, , \\ & &  \\
 f_3=\frac{\langle 4589\rangle  \langle 5679\rangle }{\langle 4569\rangle  \langle 5789\rangle } \, , ~~ &
 f_4=\frac{\langle 2349\rangle  \langle 3456\rangle  \langle 4589\rangle }{\langle 2345\rangle  \langle 3489\rangle  \langle 4569\rangle } \, , ~~ &
 f_5=\frac{\langle 3489\rangle  \langle 4569\rangle  \langle 5789\rangle }{\langle 3459\rangle  \langle 4789\rangle  \langle 5689\rangle } \, , \\ & &  \\
 f_6=\frac{\langle 4789\rangle  \langle 5689\rangle }{\langle 4589\rangle  \langle 6789\rangle } \, , ~~ &
 f_7=\frac{\langle 1239\rangle  \langle 2345\rangle  \langle 3489\rangle }{\langle 1234\rangle  \langle 2389\rangle  \langle 3459\rangle } \, , ~~ &
 f_8=\frac{\langle 2389\rangle  \langle 3459\rangle  \langle 4789\rangle }{\langle 2349\rangle  \langle 3789\rangle  \langle 4589\rangle } \, , \\ & &  \\
 f_9=\frac{\langle 3789\rangle  \langle 4589\rangle }{\langle 3489\rangle  \langle 5789\rangle } \, , ~~ &
 f_{10}=\frac{\langle 1289\rangle  \langle 2349\rangle  \langle 3789\rangle }{\langle 1239\rangle  \langle 2789\rangle  \langle 3489\rangle } \, , ~~ &
 f_{11}=\frac{\langle 2789\rangle  \langle 3489\rangle }{\langle 2389\rangle  \langle 4789\rangle } \, , \\ & &  \\
 f_{12}=\frac{\langle 1789\rangle  \langle 2389\rangle }{\langle 1289\rangle  \langle 3789\rangle } \, , ~~ &
 f_{13}=\frac{\langle 2789\rangle }{\langle 1789\rangle } \, , ~~ &
 f_{14}=\frac{\langle 3789\rangle }{\langle 2789\rangle } \, , \\ & &  \\
 f_{15}=\frac{\langle 4789\rangle }{\langle 3789\rangle } \, , ~~ &
 f_{16}=\frac{\langle 5789\rangle }{\langle 4789\rangle } \, , ~~ &
 f_{17}=-\frac{\langle 5679\rangle }{\langle 5678\rangle } \, ,  \\ & &  \\
 f_{18}=-\frac{\langle 5689\rangle }{\langle 5679\rangle } \, , ~~ &
 f_{19}=-\frac{\langle 5789\rangle }{\langle 5689\rangle } \, , ~~ &
 f_{20}=\frac{\langle 6789\rangle }{\langle 5789\rangle } \, . 
\end{array}
\end{align}

We find that mutation sequences of up to length 8 are sufficient to generate the entire $n=9$ rational symbol alphabet (mod cyclic rotation of the external labels). In particular, the 15 clusters reached from the initial
quiver described above by the mutation sequences
\begin{align} \label{minmutsn=9}
\begin{split}
\{&\{1, 3, 2, 5, 8, 7, 11, 12\} \, ,  ~ \{1, 5, 2, 10, 8, 10, 12, 11\} \, , ~ \{1, 5, 3, 9, 5, 8, 11, 12\} \, ,  \\
&\{2, 4, 6, 5, 9, 8, 11, 9\} \, , ~ \{2, 4, 6, 9, 5, 8, 12, 10\} \, , ~ \{2, 4, 7, 8, 11, 8, 12, 10\} \, , \\
&\{3, 1, 6, 5, 8, 9, 11, 12\} \, , \{3, 4, 2, 5, 8, 4, 7, 10\} \, , ~ \{4, 2, 8, 9, 8, 12, 10, 11\} \, , \\
&\{5, 6, 3, 7, 11, 10, 8, 12\} \, , ~ \{9, 4, 2, 5, 1, 3, 2\} \, , ~ \{9, 11, 6, 4, 8, 7, 10\} \, , \\
&\{10, 7,  5, 3, 2, 4, 5\} \, , ~  \{11, 6, 3, 2, 4, 7, 10\} \, , ~ \{12, 10, 1, 2, 4, 8, 5\}\} \, ,
\end{split}
\end{align}
suffice to generate the entire 531-letter $n=9$ rational symbol alphabet (mod cyclic rotations of the external labels). These 15 clusters contain only letters from this symbol alphabet. Again we note that
this set of minimal length mutation sequences is not unique, and that cluster
variables outside the 531-letter alphabet may be encountered at intermediate steps along
these sequences.

\section{Algebraic Letters}

In this section we show how to obtain
the algebraic letters of the the $n=9$ two-loop NMHV symbol alphabet~\cite{He:2020vob} by solving
$CZ=0$ for plabic parameterizations of non-top cells of $\Gr_+(4,9)$.  This
generalizes the corresponding analysis for $n=8$ carried
out in~\cite{Mago:2020kmp, He:2020uhb}.

\subsection{\texorpdfstring{$n=9$ Two-loop NMHV Algebraic Symbol Letters}{n=9 Two-Loop NMHV Algebraic Symbol Letters}}

In~\cite{He:2020vob} it was found that 99 multiplicatively independent algebraic symbol letters appear in the symbol of the two-loop nine-particle NMHV amplitude.  All algebraic letters of two-loop NMHV amplitudes trace their
origin to the one-loop four-mass box integral.  Here we recall some definitions useful for expressing these letters:
\begin{align}
\begin{split}
u_{abcd} &\equiv \frac{\langle a{-}1 ~ a ~ b{-}1 ~ b \rangle \langle c{-}1 ~ c ~ d{-}1 ~ d \rangle}{\langle a{-}1 ~ a ~ c{-}1 ~ c \rangle \langle b{-}1 ~ b ~ d{-}1 ~ d \rangle } \, , ~~~  v_{abcd} \equiv \frac{\langle b{-}1 ~ b ~ c{-}1 ~ c \rangle \langle  a{-}1 ~ a ~ d{-}1 ~ d\rangle}{\langle a{-}1 ~ a ~ c{-}1 ~ c \rangle \langle b{-}1 ~ b ~ d{-}1 ~ d \rangle } \, ,  \\\Delta_{abcd} &\equiv \sqrt{(1- u_{abcd} - v_{abcd})^2 - 4 u_{abcd} v_{abcd}} \, ,  \\
z_{abcd} &\equiv \frac{1}{2}( 1+ u_{abcd} - v_{abcd} + \Delta_{abcd} ) \, , ~~~ \bar{z}_{abcd} \equiv \frac{1}{2}( 1+ u_{abcd} - v_{abcd} - \Delta_{abcd} )  \,  .
\end{split}
\end{align}
We will also define 
\begin{align}
x_{abcd}^a = \frac{\langle \bar{d}, (c{-}1 , c) \cap (a,b{-}1, b ) \rangle}{\langle \bar{d} , a\rangle \langle b{-}1, b, c{-}1, c \rangle }\, , \\
x_{abcd}^b = \frac{\langle \bar{d}, (c{-}1 , c) \cap (a{-}1,a , b ) \rangle}{\langle \bar{d} , (a{-}1, a) \cap (b, c{-}1, c) \rangle  }\, , \\
x_{abcd}^c = \frac{\langle \bar{d} , c \rangle \langle a{-}1, a , b{-}1, b \rangle }{\langle \bar{d}, (a{-}1 , a) \cap (b{-}1 ,b, c ) \rangle }\, , 
\end{align}
where $x_{abcd}^{b-1}$, $x_{abcd}^{c-1}$ differ by exchanging $a \leftrightarrow a{-}1$ when the superscript is $a{-}1$, exchanging $b \leftrightarrow b{-}1$ when the superscript is $b{-}1$, and so on. With this, we can define two classes of algebraic symbol letters
\begin{align} \label{symbollettersalg1}
\mathcal{X}_{abcd}^\star \equiv \frac{(x_{abcd}^\star+1)^{-1} -\bar{z}_{dabc}}{(x_{abcd}^\star +1)^{-1} - z_{dabc}} \, , ~~~ \tilde{\mathcal{X}}_{abcd}^\star \equiv \frac{(x_{abc(d-1)}^\star+1)^{-1} - z_{dabc}}{(x_{abc(d-1)}^\star +1)^{-1} - \bar{z}_{dabc}} \, , 
\end{align}
where the star $\star$ corresponds to the six choices $a{-}1, a, b{-}1, b, c{-}1, c$ of the superscript of $x_{abcd}^\star$. We note that $\mathcal{X}^\star_{abcd}$, $\mathcal{X}^\star_{bcda}$, $\mathcal{X}^\star_{cdab}$ and $\mathcal{X}^\star_{dabc}$ all depend on the same square root $\Delta_{abcd}$. With this, we have a total of $4 \times 2 \times 6 = 48$ algebraic letters depending on each $\Delta_{abcd}$ from $\mathcal{X}_{abcd}^\star$ and $\tilde{\mathcal{X}}_{abcd}^\star$. In addition to these letters, there are two more letters depending on $\Delta_{abcd}$
\begin{align} \label{symbollettersalg2}
X_{abcd} = \frac{z_{abcd}}{\bar{z}_{abcd}}\, , ~~~ \text{and} ~~~ \tilde{X}_{abcd} = \frac{1- z_{abcd}}{1-\bar{z}_{abcd}} \, ,
\end{align}
bringing us to a grand total of $50$ algebraic letters depending on $\Delta_{abcd}$ in the most general case. However, in cases where $0 \leq m \leq 4$ of the corners of the four-mass box (from which these letters originate) contain only two particles, the number of independent letters containing $\Delta_{abcd}$, is reduced to $50 - 2m$. In addition, there are 33 multiplicative relations between the algebraic symbol letters of \eqref{symbollettersalg1} and \eqref{symbollettersalg2}, meaning that the number of independent letters containing $\Delta_{abcd}$ is reduced to $17 - 2m$. Thus in the nine-particle case, where we always have four-mass boxes with three corners containing two particles and one containing three, we have $m=3$ and thus $17-6 = 11$ letters for each $\Delta_{abcd}$. There are nine different square roots at $n=9$, so there are in total $11 \times 9 = 99$ independent algebraic symbol letters at $n=9$.

\subsection{\texorpdfstring{$n=9$ Algebraic Letters from Plabic Graphs}{n=9 Algebraic Letters from Plabic Graphs}}

\begin{figure}[ht]
\centering 
\includegraphics[width=0.65\textwidth]{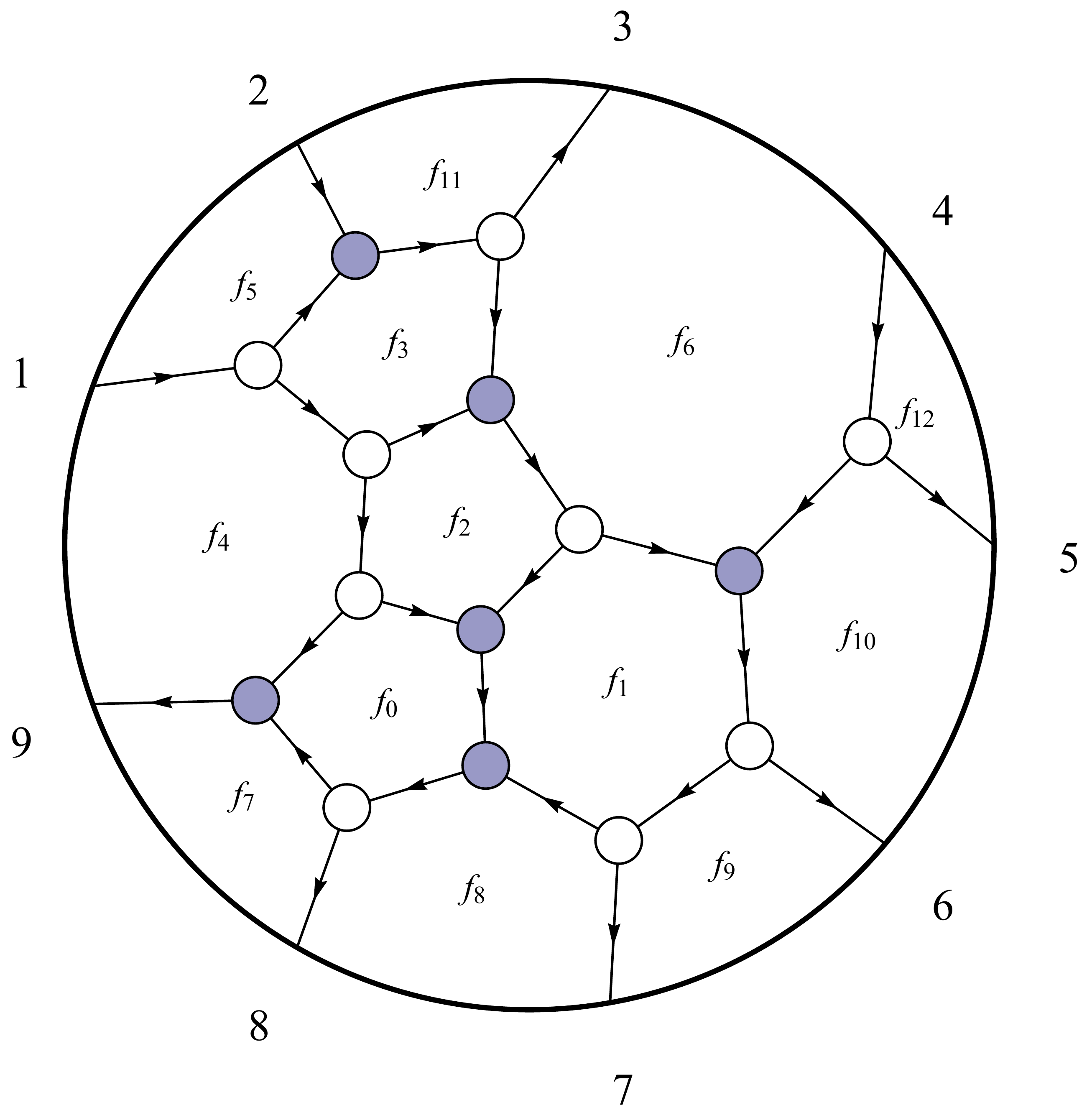}
\caption{Plabic graph associated to the
decorated permutation  $\{3, 6, 8, 5, 9, 7, 11, 10, 13\}$ in G${}_+ (3,9)$.}
\label{fig:n=9_algebraic}
\end{figure}

At $n=9$, there are two cyclic classes of positroid cells with intersection number 2  and dimension $4k$.
We recall from~\cite{Mago:2020kmp} that the latter condition is necessary for $C Z = 0$ to admit
solutions for generic $Z$, and the former condition is necessary for the solution to involve
algebraic functions (and specifically, square roots).
These two classes of cells are represented by the decorated permutations
\begin{align}
\{2, 6, 5, 8, 7, 10, 9, 13, 12\} \, , ~~ \text{and} ~~  \{2, 6, 4, 8, 7, 10, 9, 12, 14\} \,   .
\end{align}
Solutions to $C Z = 0$ from the first class of positroid cells above do not yield square roots of the type found in~\eqref{symbollettersalg1} and~\eqref{symbollettersalg2}, so we focus on the second class. In the second  class we find it computationally convenient to work with the cyclic representative $\{3, 6, 8, 5, 9, 7, 11, 10, 13\}$, which is associated to the plabic graph shown in Fig.~\ref{fig:n=9_algebraic}.

Solving $C Z = 0$, and picking one of the two solutions (the other is obtained by conjugating all roots), yields the following result in terms of the algebraic letters given in~\eqref{symbollettersalg1} and~\eqref{symbollettersalg2}:
\begin{align}
\begin{array}{ll}
f_0=\sqrt{\frac{  \langle 1239\rangle ^2 \langle  8 (23)(45)(67) \rangle \tilde{\mathcal{X}}_{9357}^a  \mathcal{X}_{5793}^c }{  \langle   2389\rangle  \langle 89(45) \cap (123) (67)\cap(123) \rangle \tilde{\mathcal{X}}_{5793}^{a-1}  \mathcal{X}_{9357}^c } } \, , ~~ &  
f_1=\sqrt{\frac{ \langle 2389\rangle  \langle  4567\rangle (\tilde{\mathcal{X}}_{5793}^{a-1})^2 \tilde{\mathcal{X}}_{7935}^{a-1}  }{  \langle 2345 \rangle \langle 6789\rangle (\mathcal{X}_{5793}^c)^2    \mathcal{X}_{7935}^c  (\tilde{\mathcal{X}}_{5793}^c)^2 }} \, , \\ & \\
f_2 =\sqrt{\frac{ \langle 1289\rangle ^2 \langle 2345\rangle  \langle   9 (23)(45)(67) \rangle \tilde{\mathcal{X}}_{9357}^{b-1} }{  \langle 1239\rangle ^2 \langle   4589\rangle  \langle 2 (45)(67)(89) \rangle \tilde{\mathcal{X}}_{9357}^a }} \, , ~~ &
f_3=\sqrt{\frac{ \langle 2389\rangle  \langle   67(23)\cap (189)  (45) \cap (189)  \rangle \tilde{\mathcal{X}}_{7935}^a \mathcal{X}_{3579}^c }{\langle 1289\rangle ^2 \langle   3 (45)(67)(89) \rangle }} \, , \\ & \\
f_4=\sqrt{\frac{  \langle   89  (45)\cap (123) (67)\cap(123)  \rangle \tilde{\mathcal{X}}_{5793}^{a-1}  \tilde{\mathcal{X}}_{7935}^{a-1} }{ \langle 2389\rangle  \langle 9 (23)(45)(67) \rangle \mathcal{X}_{5793}^c  \mathcal{X}_{7935}^c    } } \, , ~~ &
f_5=\sqrt{\frac{\langle 2389\rangle  \langle 2 (45)(67) (89) \rangle \tilde{\mathcal{X}}_{9357}^b   }{  \langle   67 (23) \cap (189) (45) \cap (189) \rangle \tilde{\mathcal{X}}_{7935}^a   \tilde{\mathcal{X}}_{9357}^{b-1}  \mathcal{X}_{3579}^c }} \, , \\ & \\
f_6=\sqrt{\frac{  \langle 4589\rangle  \langle   2 (45)(67)(89) \rangle \tilde{\mathcal{X}}_{5793}^c  \mathcal{X}_{7935}^c  \mathcal{X}_{5793}^c }{ \langle 2389\rangle  \langle 5 (23)(67)(89) \rangle \tilde{\mathcal{X}}_{5793}^{a-1}  \tilde{\mathcal{X}}_{7935}^{a-1}   \tilde{\mathcal{X}}_{9357}^{b-1} }}  \, , ~~ &
f_7=\sqrt{\frac{  \langle 9(23)(45)(67) \rangle  \mathcal{X}_{7935}^c  \mathcal{X}_{9357}^c }{  \langle 8 (23)(45)(67) \rangle \tilde{\mathcal{X}}_{7935}^{a-1}  \tilde{\mathcal{X}}_{9357}^a   } }\, , \\ & \\
f_8=\sqrt{\frac{\tilde{  \langle   4589\rangle  \langle 6 (23)(45)(89) \rangle \mathcal{X}}_{9357}^a  \mathcal{X}_{5793}^c  \mathcal{X}_{7935}^c  \tilde{\mathcal{X}}_{5793}^c }{ \langle 4567\rangle  \langle 9 (23)(45)(67) \rangle \tilde{\mathcal{X}}_{5793}^{a-1}    \tilde{\mathcal{X}}_{9357}^b }}  \, , ~~ &
f_9=\sqrt{\frac{  \langle 7 (23)(45)(89) \rangle \tilde{\mathcal{X}}_{7935}^a  }{ \langle 6(23)(45)(89) \rangle \tilde{\mathcal{X}}_{7935}^{a-1}  }} \, , \\ & \\
f_{10}=\sqrt{\frac{ \langle 6789\rangle  \langle   4 (23)(67)(89) \rangle \tilde{\mathcal{X}}_{7935}^{a-1}  \tilde{\mathcal{X}}_{9357}^b  }{  \langle 4589\rangle  \langle   7 (23)(45)(89) \rangle \tilde{\mathcal{X}}_{7935}^a  \tilde{\mathcal{X}}_{9357}    \mathcal{X}_{5793}^c  \mathcal{X}_{7935}^c}} \, , ~~ &
 f_{11}=\sqrt{\frac{  \langle 3 (45)(67)(89) \rangle  \tilde{\mathcal{X}}_{9357}^{b-1}  }{ \langle 2 (45)(67)(89) \rangle \tilde{\mathcal{X}}_{9357}^b  }} \, , \\ & \\
f_{12}=\sqrt{\frac{ \langle 5 (23)(67)(89) \rangle  \tilde{\mathcal{X}}_{9357}^a \mathcal{X}_{5793}^c  }{\langle 4 (23)(67)(89) \rangle }} \, . ~~ &
\end{array}
\label{eq:12faces}
\end{align}
All of these involve the common square root $\sqrt{\Delta_{3579}}$ of four-mass box type. The other 8 square roots can be obtained by cyclic rotations of the external labels. The 13 face variables can be expressed in terms of a basis of 11 multiplicatively independent algebraic letters:
\begin{align}
\{ \mathcal{X}_{3579}^c \, , \mathcal{X}_{5793}^c \, , \mathcal{X}_{7935}^c  \, ,  \mathcal{X}_{9357}^c \, ,  \tilde{\mathcal{X}}_{5793}^{a-1} \,  , \tilde{\mathcal{X}}_{5793}^c\,  ,  \tilde{\mathcal{X}}_{7935}^{a-1}\,  , \tilde{\mathcal{X}}_{7935}^a\, , \tilde{\mathcal{X}}_{9357}^a \, ,\tilde{\mathcal{X}}_{9357}^{b-1} \, , \tilde{\mathcal{X}}_{9357}^b  \} \, .
\end{align}

Performing all possible mutations on the internal faces of the plabic graph in Fig.~\ref{fig:n=9_algebraic}, we find an additional 12 unique factors, which can be expressed in terms of~\eqref{eq:12faces} as
\begin{align}
\begin{split}
&\begin{array}{lll}
1 + f_0 \, , ~~ & 1 + f_1  \, , ~~ & 1+ f_2 \, , 
\\ & & \\
1+ f_3 \, , ~~ & 1+ (1+f_2) f_0 \, , ~~ &  1 + (1+f_1) f_2 \, , 
\\ & & \\
1+ f_2 (1+f_3) \, , ~~ & 1+ f_0 + f_0 f_2 (1+f_1) \, , ~~ & 1+ f_0 + f_0 f_2 (1+f_3) \, , 
\end{array}\\ \\
&\hspace{2pt}1 + (1 + f_1) f_2 (1 + f_3) \, ,  ~~~~~  1+ f_0 + f_0 (1+f_1 ) f_2 (1+f_3) \, ,  \\
\\
&\hspace{2pt}1+ f_2 + f_0 (1+(1+f_1) f_2) (1+ f_2 (1+f_3)) \, .
\end{split}
\end{align}
Altogether, we therefore encounter a total of 25 algebraic factors associated to this cell of $\Gr_+(3,9)$. We find that 20 of these 25 factors are multiplicatively independent; including, of course, the 11 $\sqrt{\Delta_{3579}}$-containing algebraic letters that appear in the 2-loop nine-particle NMHV amplitude.  The additional algebraic letters that we find may appear in higher, not yet computed nine-particle amplitudes, or they may be analogs of the ``non-cluster variable'' rational quantities that generally appear when solving $C Z = 0$ for non-top cells (see~\cite{Mago:2020kmp, He:2020uhb} for examples).\\

\acknowledgments

We are grateful to L.~Ren for collaboration on the related work~\cite{RSV} and to N.~Arkani-Hamed for numerous encouraging 
conversations.
This work was supported in part by the US Department of Energy under contract
{DE}-{SC}0010010 Task A and by Simons Investigator Award \#376208 (AV).
Plabic graphs were drawn with the help of~\cite{Bourjaily:2012gy}.

\end{document}